\documentclass[aps,prd,a4paper,preprintnumbers,superscriptaddress,twocolumn,floatfix,showpacs,nofootinbib]{revtex4}
\usepackage{amssymb}
\usepackage{amsmath}

\usepackage{float,graphicx}

\begin{document}

\title{Varying-Alpha Cosmologies with Potentials}
\author{John~D.~Barrow}
\email[Email address: ]{j.d.barrow@damtp.cam.ac.uk}
\affiliation{DAMTP, Centre for Mathematical Sciences, University of Cambridge, Cambridge
CB3 0WA, United Kingdom}
\author{Baojiu~Li}
\email[Email address: ]{b.li@damtp.cam.ac.uk}
\affiliation{DAMTP, Centre for Mathematical Sciences, University of Cambridge, Cambridge
CB3 0WA, United Kingdom}

\begin{abstract}
We generalize the Bekenstein-Sandvik-Barrow-Magueijo (BSBM) model for the
variation of the fine structure 'constant', $\alpha ,$ to include an
exponential or inverse power-law self-potential for the scalar field $%
\varphi $ which drives the time variation of $\alpha $, and consider the
dynamics of $\varphi $ in such models. We find solutions for the evolution
of $\varphi $ or $\alpha $ in matter-, radiation- and dark-energy-dominated
cosmic eras. In general, the evolution of $\varphi $ is well determined
solely by either the self-potential or the coupling to matter, depending on
the model parameters. The results are general and applicable to other models
where the evolution of a scalar field is governed by a matter coupling and a
self-potential. We find that the existing astronomical data stringently
constrains the possible evolution of $\alpha $ between redshifts $z\simeq
1-3.5$ and the present, and this leads to very strong limit on the allowed
deviation of the potential from that of a pure cosmological constant.
\end{abstract}

\pacs{98.80.Es, 98.80.Bp, 98.80.Cq}

\maketitle

\section{Introduction}

\label{sect:Introduction}

For the first time there is a body of detailed astronomical
evidence consistent with the time variation of a traditional
constant of Nature. The observational programme of Webb \emph{et
al}. \cite{webb1,webb2} has completed detailed analyses of 128
Keck-HIRES quasar absorption line systems at redshifts $0.5<z<3$
using the many-multiplet method to compare separations between
line separations affected by special-relativistic effects and
found evidence consistent with the fine structure constant at
redshift $z$, $\alpha (z)$, having been \textit{smaller} in the
past, at redshifts $z\simeq 1-3.5.$ The shift in the value of
$\alpha $ between its value $\alpha (z)$ at redshift $z$ and its
present-day value $\alpha (0)$, for all the data is given
provisionally by
\begin{equation*}
\Delta \alpha /\alpha \equiv \lbrack \alpha (z)-\alpha (0)]/\alpha
(0)=(-0.57\pm 0.10)\times 10^{-5}.
\end{equation*}
Subsequent reduction of another data set of 23 VLT-UVES quasar absorption
systems $0.4\leq z\leq 2.3$ by Chand \emph{et al}., \cite{chand1,chand2}
using a partial version of the many-multiplet method at first produced a
result consistent with no variation in $\alpha $, with an unusually small
uncertainty, $\Delta \alpha /\alpha =(-0.06\pm 0.06)\times 10^{-5}.$
However, the data reduction did not allow $\Delta \alpha $ to be a free
parameter in the data fitting, and a reanalysis of the same data set by
Murphy \emph{et al}. \cite{murph} using the full many-multiplet method
increases the uncertainties sixfold, and leads to a revised bound of%
\begin{equation*}
\Delta \alpha /\alpha =(-0.64\pm 0.36)\times 10^{-5}.
\end{equation*}
Any present-day variation of $\alpha $ can also be constrained by direct
laboratory comparisons of clocks based on different atomic frequency
standards over a period of months or years. Until recently, the most
stringent atomic clock constraints on any current temporal variation of $%
\alpha $ were
\begin{equation*}
\dot{\alpha}/\alpha =(-3.3\pm 3.0)\times 10^{-16}\,yr^{-1},
\end{equation*}%
which arose by combining measurements of the frequencies of Sr
\cite{blatt}, Hg \cite{fortier}, Yb \cite{peiknew}, and H
\cite{fischer} relative to Caesium; Cing\"{o}z \emph{et al}.
\cite{cingoz} have also recently reported a less stringent limit
of $\dot{\alpha}/\alpha =-(2.7\pm 2.6)\times 10^{-15}\,yr^{-1}$.
If the systematic errors can be fully understood, an ultimate
sensitivity of $10^{-18}\,yr^{-1}$ may be possible with this
method \cite{nguyen}. If a linear variation in $\alpha $ is
assumed then the Murphy \emph{et~al}. quasar measurements equate
to $\dot{\alpha}/\alpha =(6.4\pm 1.4)\times 10^{-16}\,yr^{-1}$
\cite{webb1, webb2}. If the variation is due to a light scalar
field described by a theory like that of Bekenstein and Sandvik,
Barrow and Magueijo (BSBM) \cite{bek, bsbm}, then the rate of
change in the constants is exponentially damped during the recent
dark-energy-dominated era of accelerated expansion, and one
typically predicts a present-day value of
\begin{equation*}
\emph{\ }\dot{\alpha}/\alpha =1.1\pm 0.3\times 10^{-16}\,yr^{-1}\emph{\ }
\end{equation*}%
by direct extrapolation from the Murphy \emph{et al}. data \cite{webb1,
webb2}. This is not ruled out by the atomic clock constraints mentioned
above. For comparison, the Oklo natural reactor constraints, which are based
on the need for the $\mathrm{Sm}^{149}+n\rightarrow \mathrm{Sm}^{147}+\gamma
$ neutron capture resonance at $97.3~\mathrm{MeV}$ to have been present $%
1.8-2~\mathrm{Gyr}$ ago at $z=0.15$, as first pointed out by Shlyakhter \cite%
{sh}, are currently \cite{fuj} $\Delta \alpha /\alpha =(-0.8\pm 1.0)\times
10^{-8}$ or $(8.8\pm 0.7)\times 10^{-8}$ (because of the double-valued
character of the neutron capture cross-section with reactor temperature) and
\cite{lam} $\Delta \alpha /\alpha >4.5\times 10^{-8}$ $(6\sigma ),$ when the
non-thermal neutron spectrum is taken into account. However, there remain
significant environmental uncertainties regarding the reactor's early
history and the relationship between changes in the resonance energy level
and those in the values of any underlying constants. For reviews of the
wider issue of varying constants in addition to $\alpha $, see the reviews
in refs.~\cite{revs}, and for some implications of the unification of
fundamental forces see refs.~\cite{unify}.

Recently, Rosenband \emph{et al}. \cite{Rosenband} measured the ratio of
aluminium and mercury single-ion optical clock frequencies, $f_{\mathrm{Al+}%
}/f_{\mathrm{Hg+}}$, at intervals over a period of about a year. From these
measurements, the linear rate of change in this ratio was found to be $%
(-5.3\pm 7.9)\times 10^{-17}\,yr^{-1}$ (but see ref. \cite{bs} for some
refinements). These measurements provides the strongest limit yet on any
temporal drift in the value of $\alpha $:
\begin{equation*}
\ \dot{\alpha}/\alpha =(-1.6\pm 2.3)\times 10^{-17}\,yr^{-1}.\
\end{equation*}%
This limit is strong enough to exclude theoretical explanations of the
change in $\alpha $ reported by Webb \emph{et al}. \cite{webb1, webb2} based
on the slow variation of an effectively massless scalar field \cite{bsbm},
even allowing for the damping by cosmological acceleration, unless there is
a significant effect that slows the locally observed effects of changing $%
\alpha $ on cosmological scales (for a detailed analysis of global-local
coupling of variations in constants (see Refs.~\cite{shawb}).

Theories in which $\alpha $ varies will in general lead to violations of the
weak equivalence principle (WEP). This is because the $\alpha $ variation is
carried by a scalar field, $\varphi ,$ and this couples differently to
different nuclei because they contain different numbers of electrically
charged particles (protons). The theory discussed here has the interesting
consequence of leading to a relative acceleration of order $10^{-13}$ \cite%
{bmswep} if the free coupling parameter is fixed to the value given in Eq.~(%
\ref{om}) using a best fit of the theories cosmological model to the quasar
observations of refs.~\cite{webb1, webb2}. Other predictions of WEP
violations have also been made in refs. \cite{poly, zal, dam}. The
observational upper bound on this parameter from direct experiment is just
an order of magnitude larger, at $10^{-12},$ and limits from the motion of
the Moon are of similar order, \cite{nord}, but space-based tests planned
for the STEP mission are expected to achieve a sensitivity of order $%
10^{-18} $ and will provide a completely independent check on theories of
time-varying $e$ and $\alpha $ \cite{wep, step}.

In view of this tension between direct local measurements and astronomical
measurements of the fine structure 'constant' it is important to explore the
widest possible range of self-consistent theoretical models for the
time-evolution of $\alpha $ so as to understand the possible evolutions of $%
\Delta \alpha /\alpha $ over the range $0<z<6$ that spans the astronomical,
geochemical, and laboratory measurements. In the remainder of this paper we
will present cosmological extensions to the simple BSBM scalar field models
for varying $\alpha $ that include a non-zero self-interaction potential, $%
V(\varphi )$ for the scalar field, $\varphi $, carrying the spacetime
evolution of $\alpha $. We will consider two representative theories, where $%
V$ has exponential and power-law variation, respectively, and determine the
solutions for the cosmological evolution and the time-variation of $\alpha $
during the radiation, dust, and dark-energy dominated eras of the universe.

The organization of this paper is as follows: in \S ~\ref{sect:varying_alpha}
and \S ~\ref{sect:cosmo_eqns} we present the theory of varying $\alpha $
based on the coupling of a scalar field to the electromagnetically charged
matter and list the relativistic equations for the investigations of this
theory. In \S ~\ref{sect:BSBM} and \S ~\ref{sect:pot_only} we review,
respectively, the cosmological evolutions of the scalar field $\varphi $ for
the model with no scalar field self-interaction potential (just coupling
with matter) and for quintessence models with exponential and inverse
power-law self-potentials. The \S ~\ref{sect:pot_and_coup} is devoted to an
investigation of how the scalar field $\varphi $ evolves if both the matter
coupling term and the bare self-potential are non-zero, which is
supplemented by the numerical examples shown in \S ~\ref{sect:numerics}.
Finally, conclusions are drawn in \S ~\ref{sect:conclusion}.

\section{BSBM scalar-field theories for varying $\protect\alpha $}

\label{sect:varying_alpha}

There are a number of possible theories allowing for the variation of the
fine structure constant, $\alpha $. In the simplest cases we take $c$ and $%
\hbar $ to be constants and attribute variations in $\alpha $ to changes in $%
e$ or the permittivity of free space (see \cite{am} for a discussion of the
meaning of this choice). This is done by letting $e$ take on the value of a
real scalar field which varies in space and time. Thus $e_{0}\rightarrow
e=e_{0}\epsilon (x^{\mu }),$ where $\epsilon $ is a dimensionless scalar
field and $e_{0}$ is a constant denoting the present value of $e$. This
operation implies that some well established assumptions, like charge
conservation, must give way \cite{land}. Nevertheless, the principles of
local gauge invariance and causality are maintained, as is the scale
invariance of the $\epsilon $ field (under a suitable choice of dynamics)
and there is no conflict with local Lorentz invariance or covariance.

The dynamics are then constructed as follows. Since $e$ is the
electromagnetic coupling, the $\epsilon $ field couples to the gauge field
as $\epsilon A_{\mu }$ in the Lagrangian and the gauge transformation which
leaves the action invariant is $\epsilon A_{\mu }\rightarrow \epsilon A_{\mu
}+\chi _{,\mu },$ rather than the usual $A_{\mu }\rightarrow A_{\mu }+\chi
_{,\mu }.$ The gauge-invariant electromagnetic field tensor is therefore
\begin{equation}
F_{\mu \nu }=\frac{1}{\epsilon }\left( (\epsilon A_{\nu })_{,\mu }-(\epsilon
A_{\mu })_{,\nu }\right) ,
\end{equation}%
which reduces to the usual form when $\epsilon $ is constant. The
electromagnetic part of the action is still
\begin{equation}
S_{em}=-\int d^{4}x\sqrt{-g}F^{\mu \nu }F_{\mu \nu }.
\end{equation}%
and the dynamics of the $\epsilon $ field are controlled by the kinetic term
\begin{equation}
S_{\epsilon }=-\frac{1}{2}\frac{\hslash }{l^{2}}\int d^{4}x\sqrt{-g}\frac{%
\epsilon _{,\mu }\epsilon ^{,\mu }}{\epsilon ^{2}},
\end{equation}%
as in dilaton theories. Here, $l$ is the characteristic length scale of the
theory, introduced for dimensional reasons. This constant length scale gives
the scale down to which the electric field around a point charge is
accurately Coulombic. The corresponding energy scale, $\hbar c/l,$ has to
lie between a few tens of MeV and the Planck scale, $\sim 10^{19}$ GeV to
avoid conflict with experiment. This generalisation of the scalar theory
proposed by Bekenstein \cite{bek} was made by Sandvik, Magueijo and Barrow
\cite{bsm1, bsm2, bsm3, bsm4} and will be referred to as the BSBM theory. It
includes the gravitational effects of $\varphi $ and gives the field
equations:
\begin{equation}
G_{\mu \nu }=8\pi G\left( T_{\mu \nu }^{\mathrm{matter}}+T_{\mu \nu
}^{\varphi }+T_{\mu \nu }^{\mathrm{em}}e^{-2\varphi }\right) .  \label{ein}
\end{equation}%
The stress tensor of the $\varphi $ field is derived from the lagrangian $%
\mathcal{L}_{\varphi }=-{\frac{\omega }{2}}\partial _{\mu }\varphi \partial
^{\mu }\varphi $ and the $\varphi $ field obeys the equation of motion
\begin{equation}
\square \varphi =\frac{2}{\omega }e^{-2\varphi }\mathcal{L}_{\mathrm{em}}
\label{boxpsi}
\end{equation}%
where we have defined the coupling constant $\omega =(c)/l^{2}$. This
constant is of order $\sim 1$ if, as in \cite{bsbm}, the energy scale is
similar to the Planck scale. It is clear that $\mathcal{L}_{\mathrm{em}}$
vanishes for a sea of pure radiation since then $\mathcal{L}_{\mathrm{em}%
}=(E^{2}-B^{2})/2=0$. We therefore expect the variation in $\alpha $ to be
driven by electrostatic and magnetostatic energy-components rather than
electromagnetic radiation and with $\hbar =c=1,$ the fine-structure
'constant' is given by%
\begin{equation*}
\alpha /\alpha _{0}\equiv e^{2}/e_{0}^{2}=\exp (2\varphi ).
\end{equation*}

The considerations raised by Duff \cite{duff} do not impact upon
well-defined varying 'constant' theories like this, even if they appear
dimensionful. The presence of a new field, like $\varphi $, always requires
a second-order energy conservation equation, like Eq.~(\ref{boxpsi}) and the
integration of this equation always leads to a new integration constant, $%
\varphi _{0}$, with the same dimensions as $\varphi $ and so the evolution
of the dimensionless quantity $\varphi /\varphi _{0}$ involves no
ambiguities under redefinitions of units.

In order to make quantitative predictions we need to know how much of the
non-relativistic matter contributes to the RHS of Eq.~(\ref{boxpsi}). This
is parametrised by $\zeta \equiv \mathcal{L}_{em}/\rho $, where $\rho $ is
the energy density, and for baryonic matter $\mathcal{L}_{em}=E^{2}/2$. For
protons and neutrons $\zeta _{p}$ and $\zeta _{n}$ can be estimated from the
electromagnetic corrections to the nucleon mass, $0.63$ MeV and $-0.13$ MeV,
respectively \cite{zal}. This correction contains the $E^{2}/2$ contribution
(always positive), but also terms of the form $j_{\mu }a^{\mu }$ (where $%
j_{\mu }$ is the quarks' current) and so cannot be used directly. Hence, we
take a representative value $\zeta _{p}\approx \zeta _{n}\sim 10^{-4}$.
Furthermore, the cosmological value of $\zeta $ (denoted $\zeta _{m}$) has
to be weighted by the fraction of matter that is non-baryonic. Hence, $\zeta
_{m}$ depends strongly on the nature of the dark matter and can take both
positive and negative values depending on which of Coulomb-energy or
magnetostatic energy dominates the dark matter of the Universe. It could be
that $\zeta _{CDM}\approx -1$ (superconducting cosmic strings, for which $%
\mathcal{L}_{em}\approx -B^{2}/2)$, or $\zeta _{CDM}\ll 1$ (neutrinos). BBN
predicts an approximate value for the baryon density of $\Omega _{B}\approx
0.03$ (where $\Omega _{\mathrm{B}}$ is the density of matter in units of the
critical density $3H^{2}/8\pi G$) with a Hubble parameter of $\ H=60$ Kms$^{%
\mathrm{-1}}$ Mpc$^{\mathrm{-1}}$, implying $\Omega _{CDM}\approx 0.3$.
Thus, depending on the nature of the dark matter, $\zeta _{m}$ can be
virtually anything between $-1$ and $+1$. The uncertainties in the
underlying quark physics and especially the constituents of the dark matter
make it difficult to impose more certain bounds on $\zeta _{m}$.

There are a number of conclusions that can be drawn from the study of the
simple BSBM models with $\zeta _{m}<0$. These models gave a good fit to the
varying $\alpha $ implied by the quasar data of refs.~\cite{webb1,webb2}.
There is just a single parameter to fit and this is given by the choice \cite%
{bsbm}
\begin{equation}
-\frac{\zeta _{m}}{\omega }=(2\pm 1)\times 10^{-4}  \label{om}
\end{equation}%
The simple solutions of the BSBM theory predict a slow (logarithmic) time
increase of $\alpha $ during the dust era of $\ k=0$ Friedmann universes.
The cosmological constant turns off the time-variation of $\alpha $ at the
redshift when the universe begins to accelerate ($z\sim 0.7$) and so there
is no conflict between the $\alpha $ variation seen in quasars at $z\sim
1-3.5$ and the limits on possible variation of $\alpha $ deduced from the
operation of the Oklo natural reactor \cite{sh, fuj} (even assuming that the
cosmological variation applies unchanged to the terrestrial environment).
The reactor operated 1.8 billion years ago at a redshift of only $z\sim 0.1$
when no significant variations were occurring in $\alpha $. The slow
logarithmic increase in $\alpha $ also means that we would not expect to
have seen any effect yet in the anisotropy of the microwave backgrounds \cite%
{bat, avelino, spatial}: the value of $\alpha $ at the last scattering
redshift, $z=1000,$ is only 0.005\% lower than its value today. Similarly,
the essentially constant evolution of $\alpha $ predicted during the
radiation era leads us to expect no measurable effects on the products of
Big Bang Nucleosynthesis (BBN) \cite{jdb} because $\alpha $ was only 0.007\%
smaller at BBN than it is today. This does not rule out the possibility that
unification effects in a more general theory might require variations in
weak and strong couplings, or their contributions to the neutron-proton mass
difference, which might produce observable differences in light-element
nucleosynthesis, and new constraints on varying $\alpha ,$ at $z\sim
10^{9}-10^{10}.$ By contrast, varying-alpha cosmologies with $\zeta >0$ lead
to bad consequences unless the scalar field driving the alpha variations is
a 'ghost' field, with negatively coupled kinetic energy, in which case there
can be interesting cosmological consequences, \cite{bkm}. The fine structure
'constant' falls rapidly at late times and the variation is such that it
comes to dominate the Friedmann equation for the cosmological dynamics. We
regard this as a signal that such models are astrophysically ruled out and
perhaps are also mathematically badly behaved.

The earlier analyses of the cosmological solutions of the BSBM equations
considered the situation in which the scalar field driving variations in $%
\alpha $ has no self-interaction potential, $V(\varphi )\equiv 0$. In this
paper, we are going to explore some of the consequences for the
time-variation of $\alpha \equiv \exp (2\varphi )$ that arise when we
introduce a non-zero potential for the scalar field driving the variations
in $\alpha $. We will consider the representative cases of the exponential
potential $V(\varphi )=V_{0}\exp (-\lambda \varphi )$ and inverse power-law
potential $V(\varphi )=V_{\ast }\varphi ^{-\gamma },$ and and classify the
new behaviours that arise for $\lambda ,\gamma \neq 0$. We note that the
cases of $\lambda ,\gamma =0$ correspond to $V=V_{0}$, which is equivalent
to the presence of a cosmological constant. The solutions for such scenarios
were found in our earlier studies and $\varphi $ relaxes quickly to a
constant asymptotic value once the expansion starts to accelerate.

\section{Cosmological equations}

\label{sect:cosmo_eqns}

In the BSBM theory, the total action of the Universe is given by
\begin{equation*}
S=\int d^{4}x\sqrt{-g}(\mathcal{L}_{\mathrm{grav}}+\mathcal{L}_{\mathrm{%
matter}}+\mathcal{L}_{\mathrm{\varphi }}+\mathcal{L}_{\mathrm{em}%
}e^{-2\varphi }).
\end{equation*}%
The universe is described by a homogeneous and isotropic Friedmann metric
with expansion scale factor $a(t)$. The Friedmann equation is given by
\begin{equation}
H^{2}=\frac{1}{3}\left[ \rho _{m}(1+|\zeta |e^{-2\varphi })+\rho
_{r}e^{-2\varphi }+\rho _{\varphi }+\rho _{\mathrm{\Lambda }}\right] ;
\end{equation}
where we assume that the universe is spatially flat; the quantities $\rho
_{m},\rho _{r},\rho _{\varphi },\rho _{\mathrm{\Lambda }}$ are the energy
densities in non-relativistic matter, relativistic matter, scalar field and
cosmological constant (so $\rho _{\mathrm{\Lambda }}$ is a constant),
respectively. We will first consider the case where the scalar field $%
\varphi $ has no potential term, and then consider the cases with
exponential and power-law potentials.

The conservation equations for matter and radiation are given as
\begin{eqnarray}
\dot{\rho}_{m}+3H\rho _{m} &=&0, \\
\dot{\rho}_{r}+4H\rho _{r} &=&2\rho _{r}\dot{\varphi},
\end{eqnarray}%
and the scalar field equation of motion is
\begin{equation}
\ddot{\varphi}+3H\dot{\varphi}+\frac{\partial V(\varphi )}{\partial \varphi }%
=\frac{2|\zeta |}{\omega }\rho _{m}e^{-2\varphi }.
\end{equation}

Taking the time derivative of Eq.~(7), and using Eqs.~(8)-(10), we get
\begin{eqnarray}
\dot{H}&=&-\frac{1}{2}\left[ \rho _{m}(1+|\zeta |e^{-2\varphi })+\frac{4}{3}%
\rho _{r}e^{-2\varphi }+\omega \dot{\varphi}^{2}\right] .
\end{eqnarray}

In a universe filled with non-relativistic matter and the scalar field, $%
\rho _{\Lambda }=\rho _{r}=0$ this equation reduces to
\begin{equation*}
2\frac{\ddot{a}}{a}+\left( \frac{\dot{a}}{a}\right) ^{2}=-\left[ \frac{1}{2}%
\dot{\varphi}^{2}-V(\varphi )\right] .
\end{equation*}

\section{The case of $\protect\zeta <0$: $V$ constant}

\label{sect:BSBM}

This was the situation analysed in the original presentation of the BSBM
theory in Refs.~\cite{bsm1, bsm2, bsm3, bsm4} and extended to include
higher-order corrections in Ref.~\cite{bmota}, small perturbations \cite{bm}%
, and a linearised potential in ref. \cite{av}. The structure of the
cosmological solutions has an expected feature. The cosmological dynamics of
the scale factor $a(t)$, controlled by the Friedmann equation, is not
influenced to leading order by the small variations in $\varphi $. However,
the cosmological variation of $a(t)$ has a significant effect on the
dynamics of $\varphi $, and hence upon the evolution of the fine structure
'constant' $\alpha (t)$. The key results for a cosmological model that
evolves through a radiation-CDM-vacuum energy-dominated sequence of three
phases are as follows:

During the radiation era in which $a=t^{1/2}$, there is an exact solution of
\begin{equation*}
(\dot{\varphi}a^{3}\dot{)}=N\exp (-2\varphi )
\end{equation*}%
where $N>0$ is a constant defined by
\begin{equation*}
N\equiv -\frac{2\varsigma }{\omega }\rho _{m}a^{3}
\end{equation*}%
given by
\begin{equation*}
\varphi =\frac{1}{2}\log (8N)+\frac{1}{4}\log (t)
\end{equation*}%
For physically realistic choices of the parameters the logarithmic term is
never significant during the radiation era of our universe and $\varphi $ is
constant then, which would be expected since $E^{2}=B^{2}$ for the radiation
equilibrium which means that $\zeta \sim (E^{2}-B^{2})/(E^{2}+B^{2})$ is
effectively zero and $\varphi $ constant.

During the matter-dominated era $a=t^{2/3}$, and there is a late-time
asymptotic series solution of the form
\begin{eqnarray}
\varphi &\sim &\frac{1}{2}\ln \left[ 2N\log (t/t_{0})\right] +Ct^{-1}-\frac{1%
}{2}\sum_{n=1}^{\infty }\frac{(n-1)!}{[\log (t/t_{0})]^{n}},  \notag \\
\varphi &\rightarrow &\frac{1}{2}\log \left[ 2N\log (t)\right] ,  \notag
\end{eqnarray}%
with $C$ and $t_{0}$ constants, so $\alpha \propto \exp (2\varphi )$ grows
slowly, as $\log (t).$

During a late-time era dominated by a constant vacuum energy density, $\rho
_{\Lambda }=3H_{0}^{2}$, with de Sitter expansion of the form $a=\exp
(H_{0}t)$ we have late-time solutions of the form
\begin{eqnarray}
\varphi &\sim &\varphi _{0}+B\exp (-3H_{0}t)-\frac{N(3H_{0}t+1)}{9H_{0}^{2}}%
\exp (-2C-3H_{0}t),  \notag  \label{asym} \\
\varphi &\rightarrow &\varphi _{0},
\end{eqnarray}%
where $\varphi _{0},B$ and $C$ are constants. This case corresponds to the
addition of a constant potential $V=V_{0}$ for the scalar field and we see
that the effect is to turn off all time variations in $\varphi $, and hence $%
\alpha $. A constant asymptote for $\varphi (t)$ also occurs for any
accelerated expansion in which $a=t^{n}$, with $n\geq 1.$Any potential of
the form
\begin{equation*}
V(\varphi )=V_{0}+U(\varphi )
\end{equation*}%
where $U$ falls off as $exp(-\mu \varphi ^{n})$ for large $\varphi $, with $%
n>1$, will result in $\varphi $ approaching a constant value as $%
t\rightarrow \infty $ so long as the kinetic energy of the field is
negligible compared to the matter density, which is the physically realistic
situation. In contrast, the kinetic energy $\frac{1}{2}\dot{\varphi}^{2}$
may dominate as $t\rightarrow 0$. If it does then it leads to evolution of
the scale factor with $a=t^{1/3}$ and an exact solution for the scalar field
evolution of the form \cite{bmota}
\begin{equation*}
\varphi =\frac{1}{2}\log \left( \frac{N}{4}\right) -\log (E)+\frac{1}{2}\log
(t)+\log \left[ (\frac{t_{0}}{t})^{E}+(\frac{t}{t_{0}})^{E}\right] ,
\end{equation*}%
where $E$ and $t_{0}$ are constants; this solution approaches $\varphi =(%
\frac{1}{2}-E)\log (t)$ as $t\rightarrow 0$ and $\varphi =(\frac{1}{2}%
+E)\log (t)$ as $t\rightarrow \infty $, so the fine structure constant
evolves as $\alpha \propto t^{1\pm 2E\mathrm{\ \ }}$in these limits if the
kinetic energy dominates.

\section{The case of $\protect\zeta =0:\protect\alpha $ constant}

\label{sect:pot_only}

When $\zeta =0$, there is no coupling of the scalar field to the
electromagnetic matter, the problem reduces to the cosmology of a scalar
field in the presence of a perfect fluid and there is no variation of the
fine structure 'constant' $\alpha $. In order to include the effects of a
self-interaction potential, we shall assume two popular choices, an
exponential potential and an inverse power-law potential for the scalar
field,
\begin{eqnarray}
V(\varphi ) &=&V_{0}\exp (-\lambda \varphi ),  \notag \\
V(\varphi ) &=&V_{\ast }\varphi ^{-\gamma }  \notag
\end{eqnarray}%
in which $\lambda $ and $\gamma $ are dimensionless constants and $V_{0}\geq
0,V_{\ast }\geq 0$ are constants with dimensions [mass]$^{4}$ and [mass]$%
^{4+\gamma }$, respectively. These two potentials have been studied
extensively; they can be used to obtain power-law inflation when the scalar
field is the only matter source, and have scaling solutions where the energy
density of the scalar field evolves in proportion to the density of the
dominant\ fluid component of the universe in the presence of matter and
radiation. With the potential added, the field equations become:
\begin{eqnarray}
3H^{2} &=&\rho _{m}+\tilde{\rho}_{r}+\frac{1}{2}\dot{\varphi}^{2}+V(\varphi
)+\rho _{\Lambda }, \\
\ddot{\varphi}+3H\dot{\varphi}+V^{\prime }(\varphi ) &=&0, \\
\dot{\rho}_{m}+3H\rho _{m} &=&0, \\
\dot{\tilde{\rho}}_{r}+4H\tilde{\rho}_{r} &=&0,
\end{eqnarray}%
in which we have defined $\tilde{\rho}_{r}\equiv \rho _{r}\exp (-2\varphi )$%
. We shall discuss two particular potentials in turn below in preparation
for the discussion of the situation where varying $\alpha $ is introduced.

\subsection{Exponential potential}

For the exponential potential it is well known \cite{Ferreira1998, cop} that
scaling solutions for $\varphi $ exist when the universe is dominated by
either radiation or matter. We summarize these solutions here and also
derive a leading-order solution for the scalar field in a universe dominated
by dark energy (the dark energy is not due to the scalar field $\varphi $
here, but to the other matter).

\subsubsection{Radiation-dominated solution}

In the radiation-dominated era, we could neglect the non-relativistic matter
species and vacuum energy density ($\rho _{m}=\rho _{\Lambda }=0$) and
obtain the following solution:
\begin{eqnarray}
a &\propto &t^{1/2}, \\
H &=&\frac{1}{2t}, \\
\tilde{\rho}_{r} &=&\tilde{\rho}_{r0}\frac{t_{0}^{2}}{t^{2}}, \\
\varphi &=&\varphi _{0}+\frac{2}{\lambda }\log \frac{t}{t_{0}},
\end{eqnarray}%
where $\tilde{\rho}_{r0},t_{0}$ and $\varphi _{0}$ are constants. It is easy
to see that the scalar field energy density is given by
\begin{equation}
\rho _{\varphi }=\frac{1}{2}\dot{\varphi}^{2}+V(\varphi )\ \equiv \ \left(
\frac{2}{\lambda ^{2}}+\tilde{V}_{0}t_{0}^{2}\right) \frac{1}{t^{2}}
\end{equation}
in which we have defined $\tilde{V}_{0}=V_{0}\exp (-\lambda \varphi_{0})$.
Thus, $\rho _{\varphi }\propto t^{-2}$ scales in proportion to the radiation
energy density $\tilde{\rho}_{r}$ and their ratio is kept constant during
the evolution.

Note the Friedmann equation and the scalar field equation of motion give two
algebraic relations between the constants defining the scaling solution:
\begin{equation}
\frac{3}{4}=\left[ \tilde{\rho}_{r0}t_{0}^{2}+\frac{2}{\lambda ^{2}}+\tilde{V%
}_{0}t_{0}^{2}\right] \text{ and }1=\lambda ^{2}\tilde{V}_{0}t_{0}^{2},
\notag
\end{equation}%
and so
\begin{equation}
\tilde{V}_{0}t_{0}^{2}=\frac{1}{\lambda ^{2}}\text{ and }\tilde{\rho}%
_{r0}t_{0}^{2}=\frac{3(\lambda ^{2}-4)}{4\lambda ^{2}}.  \notag
\end{equation}%
Hence, the constant fractional energy densities are given by
\begin{eqnarray}
\Omega _{r} &=&\frac{\lambda ^{2}-4}{\lambda ^{2}}, \\
\Omega _{\varphi } &=&\frac{4}{\lambda ^{2}}.
\end{eqnarray}

\subsubsection{Matter-dominated solution}

Similarly, in the matter-dominated era we can neglect the radiation and
vacuum densities to obtain the following solution:
\begin{eqnarray}
a &\propto &t^{2/3}, \\
H &=&\frac{2}{3t}, \\
\rho _{m} &=&\rho _{m0}\frac{t_{0}^{2}}{t^{2}}, \\
\varphi &=&\varphi _{0}+\frac{2}{\lambda }\log \frac{t}{t_{0}},
\end{eqnarray}%
where $\rho _{m0},t_{0}$ and $\varphi _{0}$ are new constants. Here,
Eq.~(21) still holds and $\rho _{\varphi }$ now scales as $\rho _{m}$. The
Friedmann equation and the scalar field equation of motion give two
algebraic relations between these quantities
\begin{equation}
\frac{4}{3}=\left[ \rho _{m0}t_{0}^{2}+\frac{2}{\lambda ^{2}}+\tilde{V}%
_{0}t_{0}^{2}\right] \text{ and }2=\lambda ^{2}\tilde{V}_{0}t_{0}^{2},
\notag
\end{equation}%
which lead to
\begin{equation}
\tilde{V}_{0}t_{0}^{2}=\frac{2}{\lambda ^{2}}\text{ and }\rho _{m0}t_{0}^{2}=%
\frac{4(\lambda ^{2}-3)}{3\lambda ^{2}},  \notag
\end{equation}%
and so the constant fractional energy densities are given by
\begin{eqnarray}
\Omega _{m} &=&\frac{\lambda ^{2}-3}{\lambda ^{2}}, \\
\Omega _{\varphi } &=&\frac{3}{\lambda ^{2}}.
\end{eqnarray}

Note that in order for $\Omega _{\varphi }<1$ during the radiation era, $%
\lambda ^{2}>4$ is required, and this then ensures $\Omega _{\varphi }<1$
during the dust era and $\Omega _{m}>0$.

\subsubsection{Dark-Energy-dominated solution}

Recent observations suggest that the universe is currently, and will remain,
dominated by a gravitationally repulsive form of matter dubbed dark energy.
In the simplest scenario, this is just a cosmological constant for which the
expansion rate of the Universe will tend to a constant, while in other
models it can be exotic matter, or changes to the law of gravitation, which
drive a different future evolution of the universe. Here we consider the
evolution of our $\varphi $ driven by the exponential potential in the
background of simple dark energy domination. For simplicity we consider just
two cases, where the cosmic expansion factor is either exponential, $a=\exp
(H_{0}t)$, $H_{0}$ constant, or a power-law in time, $a\propto t^{\beta }$
with $\beta >1$.

\paragraph{Case 1}

This is a cosmological constant ($\Lambda $) dominated universe, for which
the scale factor evolves as $a\propto \exp (\varsigma t)$ where $\varsigma
\equiv \sqrt{\Lambda /3}$ and so the EOM for $\varphi $ becomes:
\begin{equation}
\ddot{\varphi}+3\varsigma \dot{\varphi}=W\exp (-\lambda \varphi )
\end{equation}%
where we have defined $W\equiv \lambda V_{0}$. This equation is very similar
to the scalar field EOM for the BSBM model in the dust-dominated era. It has
no closed analytical solution and we shall seek a self-consistent
approximate solution following the logic in ref. \cite{bsm1}. We start from
the ansatz that in the $\Lambda $ dominated era the field $\varphi $ is
slowly-rolling, and it is easy to obtain the slow-roll solution $\varphi
\sim \frac{1}{\lambda }\log (\lambda Wt/3\varsigma )$ by setting $\ddot{%
\varphi}=0$. Let us next make the following approximation by an asymptotic
series:
\begin{equation*}
\varphi =\frac{1}{\lambda }\log \left[ \frac{\lambda W}{3\varsigma }t\right]
+\sum_{n=1}^{\infty }a_{n}t^{-n}
\end{equation*}%
where $a_{n}$ are some constant coefficients. Substituting this back into
the scalar field EOM Eq.~(30) we have
\begin{eqnarray}
&&-\frac{1}{\lambda t^{2}}+\sum_{n=1}^{\infty }n(n+1)\frac{a_{n}}{t^{n+2}}+%
\frac{3\varsigma }{\lambda t}-\sum_{n=1}^{\infty }n\frac{a_{n}}{t^{n+1}}
\notag \\
&=&\frac{3\varsigma }{\lambda t}\exp \left[ -\lambda \sum_{n=1}^{\infty }%
\frac{a_{n}}{t^{n}}\right] \rightarrow \frac{3\varsigma }{\lambda t},  \notag
\end{eqnarray}%
as $t\rightarrow \infty $. Choose appropriate $a_{n}$ so that the terms $%
1/t^{r}$ with $r\geq 2$ cancel, we find that the solution for $\varphi $ can
be written as
\begin{equation}
\varphi =\frac{1}{\lambda }\log \left[ \frac{\lambda W}{3\varsigma }t\right]
-\frac{1}{2}\left[ \frac{1}{t}+\frac{1}{t^{2}}+\frac{2}{t^{3}}+\cdots +\frac{%
(r-1)!}{t^{r}}+\cdots \right] .
\end{equation}%
It is clear that as time grows, the asymptotic series becomes less important
and so the slow-roll solution is ever improved. Eq.~(31) is a good
approximation when $t$ is large as we have seen in the derivation; when $t$
is small, Eq.~(30) could be linearized as
\begin{equation*}
\ddot{\varphi}+3\varsigma \dot{\varphi}=W
\end{equation*}%
where we assume the initial value of $\varphi $ is zero. The solution is
then
\begin{equation}
\varphi =\varphi _{c}+A\exp (-3\varsigma t)+\frac{Wt}{3\varsigma }\
\rightarrow \ \varphi _{c}+\frac{Wt}{3\varsigma }
\end{equation}%
where $\varphi _{c}$ and $A$ are constants of integration. The linear term
in $t$ seems to be the second Taylor term of the slow-roll solution. We can
see that in both solutions the scalar field $\varphi $ will not tend to
constant when $t$ goes large, which is in contrast to the case of BSBM.

\paragraph{Case 2}

This is described by $a\propto t^{n}$ ($n>1$), for which the scalar field
EOM could be written as
\begin{equation}
\ddot{\varphi}+\frac{3n}{t}\dot{\varphi}=W\exp (-\lambda \varphi )
\end{equation}
which has an exact solution
\begin{equation}
\varphi =\frac{1}{\lambda }\log \frac{\lambda W}{2(3n-1)}+\frac{2}{\lambda}%
\log t.
\end{equation}
Again, we find a logarithmic behaviour of $\varphi $ in the acceleration
era, which means that $\varphi $ will never approach a constant. This is, of
course, not surprising because we know that the exponential potential has
tracking behaviour for any power-law background expansion with $n>1/3$, no
matter whether it is $n<1$ (matter and radiation dominations) or $n>1$ (dark
energy domination).

Of course, we also need to justify the assumption that the energy density in
the scalar field is always subdominant. In case 1 this is obvious because
the $\Lambda $ has a constant energy density while the scalar field has a $%
p/\rho $ ratio which is greater than $-1$, meaning that its energy density
decays continuously. For case 2 we have $\rho _{\varphi }\propto
t^{-2}\propto \rho _{\mathrm{DE}}$ and it again exactly tracks the dominant
component in the universe. In both cases there is no way for $\varphi $ to
come to dominate the total energy density.

\subsection{Inverse power-law potential}

We turn next to the inverse power-law potential $V=V_{\ast }\varphi
^{-\gamma },$ with $\gamma $ a positive constant. It is well known \cite%
{Invpow} that this potential also permits tracking behaviour of the scalar
field $\varphi $.

\subsubsection{Radiation- and Matter-dominated solutions}

Suppose the background universe expands according to $a\propto t^{n}$ and
the energy density in $\varphi $ is only subdominant, then $\varphi $ has
the solution:
\begin{equation}
\varphi =At^{\frac{2}{\gamma +2}}
\end{equation}%
where $A$ is constant to be fixed. To determine the value of $A$, take the
time derivatives of $\varphi $
\begin{eqnarray}
\dot{\varphi} &=&\frac{2A}{\gamma +2}t^{-\frac{\gamma }{\gamma +2}},  \notag
\\
\ddot{\varphi} &=&-\frac{2A\gamma }{(\gamma +2)^{2}}t^{-\frac{2\gamma +2}{%
\gamma +2}}  \notag
\end{eqnarray}%
and insert them together with $H=n/t$ into the scalar field EOM, we get an
algebraic equation for $A$ which has the solution
\begin{equation}
A=\left[ \frac{\gamma (\gamma +2)^{2}V_{\ast }}{6n(\gamma +2)-2\gamma }%
\right] ^{\frac{1}{\gamma +2}}.
\end{equation}

Since $\varphi \propto t^{\frac{1}{\gamma +2}}$, it is easy to see that $%
\rho _{\varphi }=\frac{1}{2}\dot{\varphi}^{2}+V(\varphi )\propto t^{-\frac{%
2\gamma }{\gamma +2}}\rightarrow t^{-2}$ for $\gamma \gg 1$. As the dominant
component in the universe also has an energy density scaling as $t^{-2}$, we
see that the energy density of $\varphi $ simply tracks the dominant
component, which is radiation in the radiation era and dust in matter era.

Note that this tracking behaviour is only approximate for $\gamma \gg 1$,
while in reality $\rho _{\varphi }$ decays slower than $\rho _{\mathrm{%
dominant}}$. This means that the fractional energy density of the scalar
field $\varphi $ is ever increasing and eventually will no longer be
subdominant. However, for enough large $\gamma $ this will take a very long
time so the issue will not bother us for some time.

\subsubsection{Dark-Energy-dominated solution}

We next consider the evolution of $\varphi $ in a dark-energy-dominated
universe. Again, we consider two cases, case 1 for $\Lambda $ domination
where $H$ is a constant and case 2 for a power-law inflation $a\propto t^{n}$
($n>1$). Obviously case 2 has the same behaviour as in the radiation or
matter-dominated universes but with the value of $n$ in Eq.~(36) changed,
and so we will not consider it again here except stating that in the $%
a\propto t^{n}$ dark energy era the field $\varphi $ does not stop growing.

In the first case, of $\Lambda $ domination, we now write the scalar field
EOM as
\begin{equation}
\ddot{\varphi}+3\varsigma \dot{\varphi}=\frac{\gamma V_{\ast }}{\varphi
^{\gamma +1}}.
\end{equation}%
The slow-roll solution to this equation is
\begin{equation}
\varphi \sim \left[ \frac{\gamma (\gamma +2)V_{\ast }}{3\varsigma }t\right]
^{\frac{1}{\gamma +2}}.
\end{equation}%
When $t$ goes large, the $\ddot{\varphi}$ term will be less and less
important because $\ddot{\varphi}/\dot{\varphi}\propto 1/t$ and so the
slow-roll solution is ever improved. Also, again the energy density in the
scalar field $\varphi $ decays in time so that its fractional energy density
always decreases and it will never dominate the total energy density.
Eq.~(38) indicates that $\varphi $ will continue to grow in the $\Lambda $%
-dominated era; however the rate of growth is lower than that in case of $%
a\propto t^{n}$ [c.f.~Eq.~(35)], and can be very low when $\gamma
\rightarrow \infty $. If needed, we could improve Eq.~(38) by adding an
asymptotic series, of which the leading terms are
\begin{equation*}
\varphi ^{\gamma +2}\sim \frac{\gamma (\gamma +2)V_{\ast }}{3\varsigma }t+%
\frac{\gamma (\gamma +1)V_{\ast }}{9\varsigma ^{2}}\log t+\mathrm{const.}+%
\mathcal{O}\left( \frac{\log t}{t}\right) .
\end{equation*}

\section{The General Case of $\protect\zeta \neq 0$ and $V \neq 0$}

\label{sect:pot_and_coup}

Our ultimate aim is to consider the evolution of the scalar field $\varphi $
when both the bare potential [c.f.~\S ~4] and the matter coupling term [c.f.~%
\S ~3] are present, from which we can learn how the fine structure
'constant' $\alpha $ evolves in time. To that end we draw the contents of
the above two sections together to get a picture of the whole evolution of $%
\varphi $ in the presence of both terms. We consider again the two cases: an
exponential potential and an inverse power-law potential.

\subsection{Exponential potential}

Before going to the general radiation and matter-dominated solutions for the
exponential potential and a matter coupling term, we first consider the
special case that arises when $\lambda =8$. The scalar field equation now
becomes Eq.~(10):
\begin{equation}
\ddot{\varphi}+3H\dot{\varphi}+\frac{\partial V(\varphi )}{\partial \varphi }%
=2\frac{|\zeta |}{\omega }\rho _{m}e^{-2\varphi }
\end{equation}%
and we still use the exponential potential given above.

We can see that Eqs.~(17 - 20) remain as in the radiation-dominated
solution. This is easy to check because for $a\propto t^{1/2}$, we have $%
\rho _{m}\propto a^{-3}\propto t^{-3/2}$, $e^{-2\varphi }=(e^{-\lambda
\varphi })^{2/\lambda }\propto t^{-4/\lambda }=t^{-1/2}$, so $\alpha $ grows
as $\exp (2\varphi )\propto t^{1/2}$, and both sides of Eq.~(39) scale as $%
t^{-2}$. In this special case the presence of the coupling term does not
influence the overall form of the solution, although Eqs.~(22-23) might be
changed (slightly) so that the (constant) fractional energy density $\Omega
_{\varphi }$ is shifted in value. For the case $\lambda =8$, originally we
had $\Omega _{\varphi }=1/16$ when $\varsigma =0$, and in the
radiation-dominated era $\rho _{m}\ll \tilde{\rho}_{r}$ so we expect the
shift to be tiny.

This discussion can be generalized to include a subsequent era dominated by
a fluid with general equation of states ($p\neq 0$). However, for a
matter-dominated era, Eqs.~(28 - 31) no longer remain an exact solution
unless $\lambda \rightarrow \infty $.

Let us now turn to the more general $\zeta \neq 0$ cases. The effective
total potential for the scalar field $\varphi $ consists of two parts, the
bare potential, $V(\varphi )$ and the electromagnetic matter couplings,
which depends on $a(t)$, via $\rho _{m}\propto a^{-3}$, so we can combine
them as
\begin{equation}
V_{eff}(\varphi )=V(\varphi )+\frac{|\zeta |}{\omega }\rho _{m}\exp
(-2\varphi ).
\end{equation}
The parameters used in this section are not specifically chosen to reproduce
the observed time variation of the fine structure constant (which we defer
to the next section), rather here we are concerned with the general dynamics
under the effective potential, which might also be useful for models of a
scalar field with self-potential coupling to dark matter.

\subsubsection{Radiation-dominated era}

No matter which of the two parts to the scalar effective potential $V_{eff%
\text{ \ }}$dominates, the field $\varphi $ will grow at most
logarithmically in the radiation era, and (except $\lambda =8$) the only
difference between the bare-potential-dominated and the coupling-dominated
solutions is the coefficient in front of $\log t$. However, if that
coefficient is very small then $\varphi $ will remain approximately constant
during the radiation era, as in BSBM with $V=0$ discussed above.

As it is the radiation era, if the scalar field makes a significant
contribution to the energy budget of the Universe then $\rho _{\varphi }\sim
V(\varphi )\gg \frac{|\zeta |}{\omega }\rho _{m}\exp (-2\varphi ),$ (notice
also that $\frac{|\zeta |}{\omega }\ll 1$ which reduces the possible
influence of the coupling terms even further), and thus $V_{eff}(\varphi
)\sim V(\varphi )$. On the other hand, if the scalar field constitutes only
a very small part of the total energy density (for example, if $\lambda \gg
1 $), then $V(\varphi )$ might be comparable to, or even much smaller than, $%
\frac{|\zeta |}{\omega }\rho _{m}\exp (-2\varphi )$.

Since the scale factor evolves as $a\propto t^{1/2}$ whichever part of the
effective potential dominates, we shall take this as the leading-order
solution to the Friedmann equation and look at the evolution of $\varphi $
under this condition. We then could rewrite the scalar field EOM as
\begin{eqnarray}
&&e^{-x}\frac{d}{dx}\left[ e^{\frac{1}{2}x}\frac{d}{dx}\varphi (x)\right]
\notag \\
&=&N\exp [-2\varphi (x)]+W\exp \left( \frac{3}{2}x\right) \exp [-\lambda
\varphi (x)]
\end{eqnarray}%
where we have defined $x\equiv \log t$, with $N\equiv 2\frac{|\zeta |}{%
\omega }\rho _{m}a^{3}$ and $W\equiv \lambda V_{0}$ constants. Clearly, the
larger $N$ is, the easier it is for the coupling term on the right hand side
to dominate, and the larger $W$ is, the easier it is for the potential term
to dominate.

\begin{figure}[tbp]
\centering \includegraphics[scale=1.23] {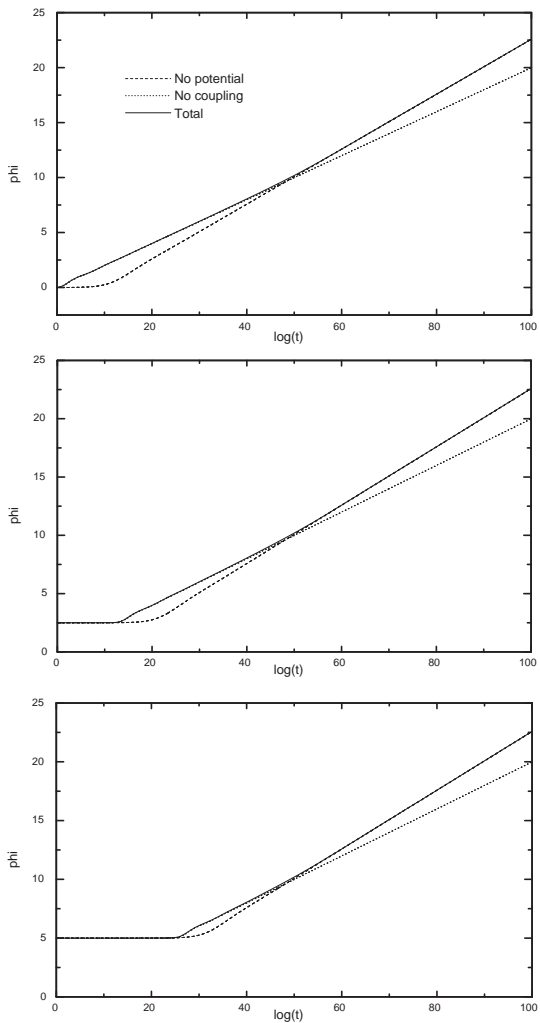} \caption{The
evolution of $\protect\varphi $ as a function of $\log t$ in a
radiation-dominated universe. The solid, dashed and dotted curves
represent the total evolution, the evolution governed solely by
the coupling term, and that governed only by the bare-potential
term, respectively. The parameters are $\protect\lambda =10$,
$N=0.001$ and $W=0.1$; the initial conditions in
the upper, middle and lower panels are $\dot{\protect\varphi}_{i}=0$ and $%
\protect\varphi _{i}=0,2.5,$and $5$ respectively.}
\label{fig:Figure1}
\end{figure}

As discussed above, we adopt a value of $\lambda >2$; the larger $\lambda $
is, the less important is the potential term for large $\varphi $. Because $%
\lambda >2$, the potential term decays faster than the coupling term, so if
the coupling term dominates at initial time, the potential term will never
become important. On the other hand, if the potential term dominates at
initial time, at some later time (if the radiation era is long enough) the
effective potential will become dominated by the coupling term and the
evolution of $\varphi $ changes accordingly.\emph{\ }

These behaviours are easy to verify numerically by solving Eq.~(41), and an
example is given in Fig.~\ref{fig:Figure1}.

\subsubsection{Matter-dominated era}

Now we turn to the matter-dominated era in which $\rho _{m}\gg \tilde{\rho}%
_{r}$ and thus the latter can be neglected. If the bare potential is the
major part of the effective potential then, according to previous analysis,
the scale factor scales as $a\propto t^{2/3}$; if the coupling term
dominates, Barrow \emph{et al.} also showed \cite{bsm1, bsm2, bsm3, bsm4,
bmota} that $a\propto t^{2/3}$ in the leading-order solution. So here we
also assume that this is a good approximation in the matter-dominated era
and look at the evolution of $\varphi $ on this background.

\begin{figure}[tbp]
\centering \includegraphics[scale=1.23] {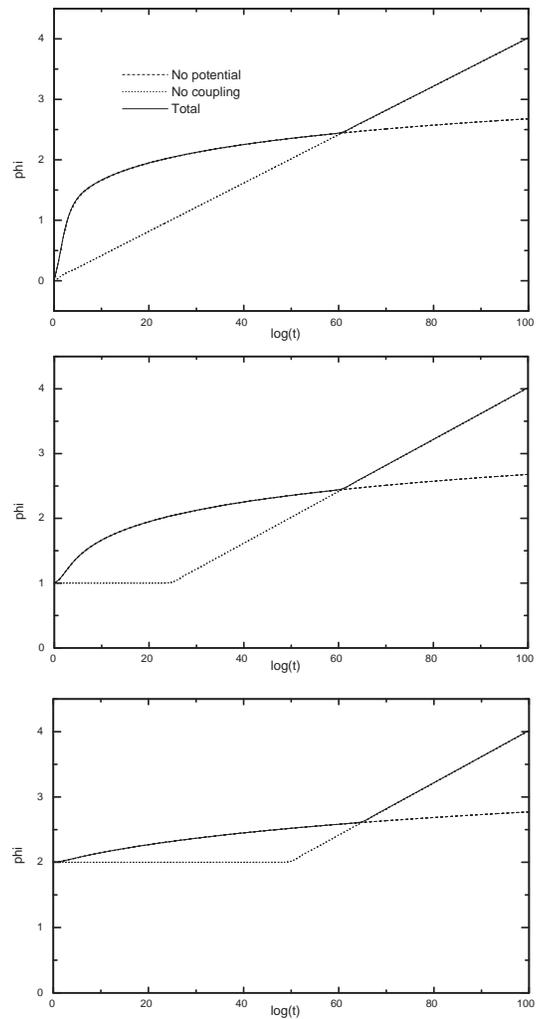} \caption{The
evolution of $\protect\varphi $ as a function of $\log t$ in a
matter-dominated background universe. The solid, dashed and dotted
curves represent  the total evolution, the evolution governed
solely by the coupling term, and that governed only by the
bare-potential term,
respectively. The parameters chosen are $\protect\lambda =50$, $N=1$ and $%
W=0.1$; the initial conditions in the upper, middle and lower panels are $%
\dot{\protect\varphi}_{i}=0$ and $\protect\varphi _{i}=0,1,$and $2$
respectively.}
\label{fig:Figure2}
\end{figure}

In this case the scalar field EOM becomes
\begin{eqnarray}
&& e^{-x}\frac{d}{dx}\left[ e^{x}\frac{d}{dx}\varphi (x)\right]  \notag \\
&=& N\exp [-2\varphi(x)]+W\exp \left(2x\right) \exp[-\lambda\varphi(x)].
\end{eqnarray}

A qualitative analysis can be made as in the case of radiation domination,
by considering the evolution without the potential or coupling term present,
respectively. In the case where only the potential term is included, we have
$\varphi \propto \log t$; if only the coupling term is presented then $%
\varphi $ evolves as $1/2\log (2N\log t)$ approximately. Now suppose that
initially the potential term dominates over the coupling term, then because
the former scales as $t^{-2}$ while the latter scales as $\rho _{m}\exp
(-2\varphi )\propto t^{-2-4/\lambda }$ and falls faster, the potential term
will become increasingly dominant and the coupling term will never become
important. On the other hand, if initially the coupling term dominates, then
the potential term and the coupling term scale as $(\log t)^{-\lambda /2}$
and $t^{-2}[\log (t)]^{-1}$, respectively, and the former always falls off
slower than the latter (however, depending on the value of $\lambda $, the
dominance of the potential term could occur very late, much later than the
transition to an acceleration era, so probably we will not see this
transition during the matter epoch). Thus, in this case, the potential term
will finally overwhelm the coupling term, and the full solution will then
track the no-coupling one. These features can also be checked numerically.
Note that increasing $N$ or $\lambda $ will help the coupling term to
dominate. The tracking is excellent in both regions. When the coupling term
dominates the fine structure 'constant' evolves as $\alpha \propto 2N\log t,$
but when the potential term dominates, $\alpha \propto (t/t_{0})^{4/\lambda
}.$

\begin{figure}[tbp]
\centering \includegraphics[scale=1.23] {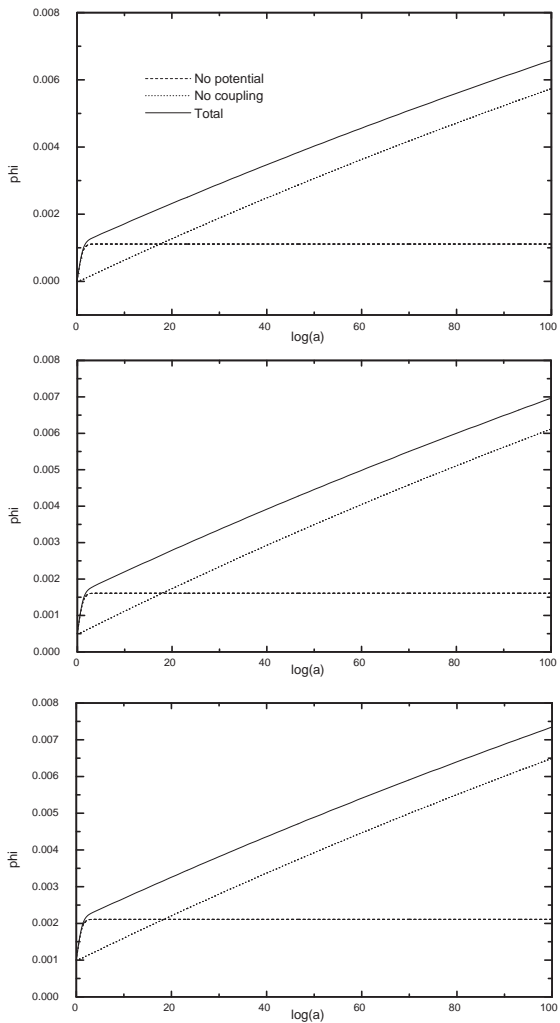}
\caption{The evolution of $\protect\varphi $ as a function of $\log a$ in a $%
\Lambda $-dominated universe. The solid, dashed and dotted curves represent
the total evolution, the evolution governed solely by the coupling term, and
that governed only by the bare-potential term, respectively. The parameters
chosen are $\protect\lambda =50$, $N=0.01$ and $W=0.0002$; the initial
conditions in the upper, middle and lower panels are $\dot{\protect\varphi}%
_{i}=0$ and $\protect\varphi _{i}=0,0.0005,$and $0.001$ respectively.}
\label{fig:Figure3}
\end{figure}

A numerical example of the behaviours discussed above is shown in Fig.~\ref%
{fig:Figure2}.

\subsubsection{Dark-energy-dominated era}

A later times the universe will become dominated by any cosmological
constant $\Lambda $ and then the scalar field EOM becomes
\begin{equation*}
\ddot{\varphi}+3\varsigma \dot{\varphi}=W\exp (-\lambda \varphi )+N\exp
(-3\varsigma t)\exp (-2\varphi ).
\end{equation*}

If the evolution is initially dominated by the bare potential term ($W$),
then we know from the above analysis that $\varphi \sim \frac{1}{\lambda }%
\log t$. So the bare-potential term in the above equation scales as $1/t$
while the coupling term scales exponentially with respect to $t,$ as $t^{-%
\frac{2}{\lambda }}\exp (-3\varsigma t)$. For large $t,$ the latter decays
faster, and the bare-potential term will eventually dominate over the
coupling term and the fine structure 'constant' will evolve all the time in
the future. A numerical example is shown in Fig.~\ref{fig:Figure3}. Note
that for this analysis we have defined a different set of parameters $%
x\equiv \log a$, with $W\equiv \lambda V_{0}/\varsigma ^{2}$ and $N\equiv
2|\zeta |\rho _{m0}/\varsigma ^{2}$ constants, so that the above EOM is
rewritten as
\begin{eqnarray}
&&\frac{d^{2}}{dx^{2}}\varphi (x)+3\frac{d}{dx}\varphi (x)  \notag \\
&=&N\exp (-3x)\exp (-2\varphi )+W\exp (-\lambda \varphi ).  \notag
\end{eqnarray}

If the dark energy drives power-law inflation $a\propto t^{n}$ ($n>1$) of
the universe today, then the analysis of the whole evolution is
qualitatively the same as for a radiation-dominated universe and will not be
repeated here.

\subsection{Inverse power-law potential}

Next, we consider the evolution of $\varphi $ under the controls of the
coupling term and a bare inverse power-law potential.

\subsubsection{Radiation-dominated era}

In the radiation-dominated era the cosmic scale factor scales as $a\propto
t^{1/2}$, while the energy densities of scalar field and dust can be
neglected. So, just as in the case of exponential potential, we can write
the scalar field EOM as
\begin{eqnarray}
&&e^{-x}\frac{d}{dx}\left[ e^{\frac{1}{2}x}\frac{d}{dx}\varphi (x)\right]
\notag \\
&=&N\exp [-2\varphi (x)]+W\exp \left( \frac{3}{2}x\right) \varphi ^{-(\gamma
+1)},
\end{eqnarray}%
where again $x=\log t,$ but now $W\equiv \gamma V_{\ast }$.

\begin{figure}[tbp]
\centering \includegraphics[scale=1.23] {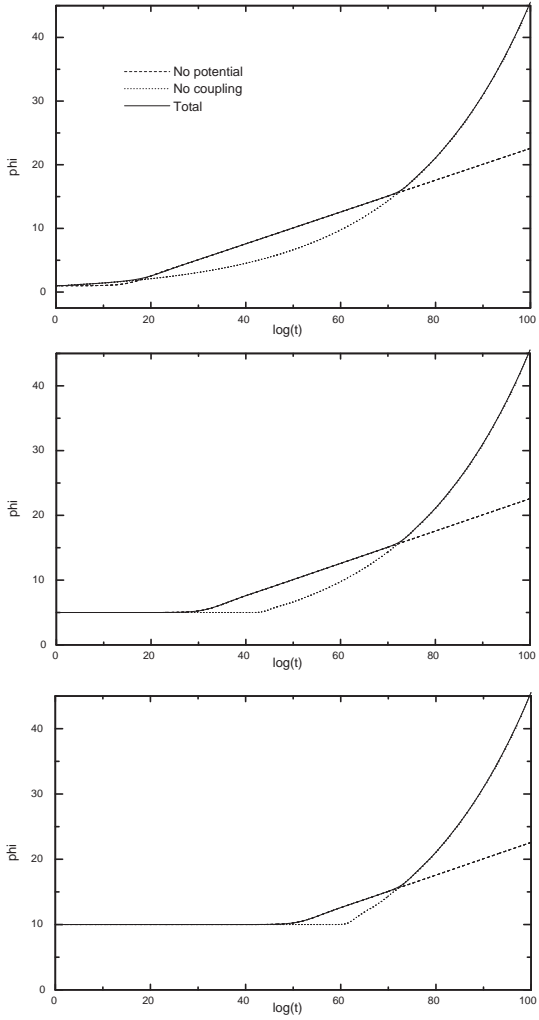}
\caption{The evolution of $\protect\varphi $ as a function of
$\log t$ in a radiation-dominated universe. The solid, dashed, and
dotted curves represent the total evolution, the evolution
governed solely by the coupling term, and that governed only by
the bare-potential term, respectively. The parameters used are
$\protect\gamma =50$, $N=0.001$ and $W=0.005$; the initial
conditions in the upper, middle and lower panels are $\dot{\protect\varphi}%
_{i}=0$ and $\protect\varphi _{i}=1,5,$ and $10$ respectively.}
\label{fig:Figure4}
\end{figure}

According the above analysis, if the bare potential term dominates then $%
\varphi $ evolves as $\varphi \sim t^{\frac{2}{\gamma +2}}\sim \exp \left(
\frac{2}{\gamma +2}x\right) $, (so $\varphi $ is exponential in $x$), while
if the coupling term dominates then $\varphi \sim \frac{1}{4}\log t\sim
\frac{1}{4}x$ (so $\varphi $ is linear in $x$). These features can be seen
clearly in the numerical example given in Fig.~\ref{fig:Figure4}. Note that
in the scalar field EOM the bare-potential term decays as $\sim \varphi
^{-(\gamma +1)}$, while the coupling term scales either as $\sim \rho
_{m}\exp (-2\varphi )\sim t^{-\frac{3}{2}}\exp (-2\varphi )\sim \varphi ^{-%
\frac{3}{4}(\gamma +2)}\exp (-2\varphi ),$ (in case the effective potential $%
V_{eff}$ is dominated by the bare potential), or as $\sim \rho _{m}\exp
(-2\varphi )\sim t^{-\frac{3}{2}}\exp (-2\varphi )\sim \exp (-8\varphi ),$
(when $V_{eff}$ is dominated by the coupling term), so obviously in both
cases when eventually $\varphi $ is large enough the bare-potential term
will dominate over the coupling term driving the evolution of $\varphi $.
This can also be seen in the figure.

\subsubsection{Matter-dominated era}

In the matter-dominated era the scale factor evolves as $a\propto t^{2/3}$,
while the energy densities of scalar field and radiation can be neglected.
In this case the scalar field EOM becomes
\begin{eqnarray}
&&e^{-x}\frac{d}{dx}\left[ e^{x}\frac{d}{dx}\varphi (x)\right]  \notag \\
&=&N\exp [-2\varphi (x)]+W\exp \left( 2x\right) \varphi ^{-(\gamma +1)},
\end{eqnarray}%
where $x$ and $W$ are as defined above.

\begin{figure}[tbp]
\centering \includegraphics[scale=1.23] {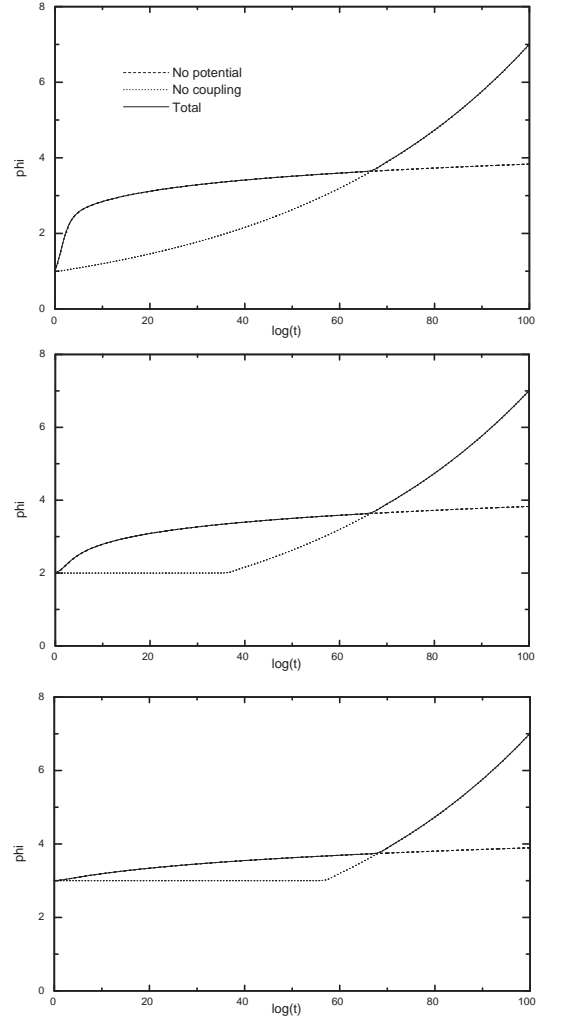}
\caption{The evolution of $\protect\varphi $ as a function of
$\log t$ in a matter-dominated universe. The solid, dashed and
dotted curves represent the total evolution, the evolution
governed solely by the coupling term, and that governed only by
the bare-potential term, respectively. The parameters chosen are
$\protect\lambda =100$, $N=10$ and $W=0.005$; the initial
conditions in the upper, middle and lower panels are $\dot{\protect\varphi}%
_{i}=0$ and $\protect\varphi _{i}=1,2,$and $3$ respectively.}
\label{fig:Figure5}
\end{figure}

From the above analysis we know that if the bare potential term dominates
then $\varphi $ evolves as $\varphi \sim t^{\frac{2}{\gamma +2}}\sim \exp
\left( \frac{2}{\gamma +2}x\right) $ (\emph{i.e.}, $\varphi $ is exponential
in $x$), while if the coupling term dominates then to the leading order $%
\varphi \sim \frac{1}{2}\log (\log t)\sim \frac{1}{2}\log x$ (\emph{i.e.}, $%
\varphi $ is logarithmic in $x$). We can also see these behaviours clearly
in the numerical results plotted in Fig.~\ref{fig:Figure5}. In the scalar
field EOM the bare-potential term decays as $\sim \varphi ^{-(\gamma +1)}$,
while the coupling term scales either as $\sim \rho _{m}\exp (-2\varphi
)\sim t^{-2}\exp (-2\varphi )\sim \varphi ^{-(\gamma +2)}\exp (-2\varphi ),$
(when $V_{eff}$ is dominated by the bare potential), or as $\sim \rho
_{m}\exp (-2\varphi )\sim t^{-2}\exp (-2\varphi )\sim \exp \left[ -2\exp
(2\varphi )\right] \exp (-2\varphi )$ (when $V_{eff}$ is dominated by the
coupling). Therefore, when $\varphi $ eventually grows large enough, the
bare-potential term will dominate over the coupling term and drive the
evolution of $\varphi $. This is verified in the numerical results, where we
can see specific tracking solutions in different epochs.

\subsubsection{Dark-energy-dominated era}

In a universe dominated by a cosmological constant, the field equation for $%
\varphi$ in the case of a power law potential can be written as
\begin{eqnarray}
\frac{d^{2}}{dx^{2}}\varphi(x) + 3\frac{d}{dx}\varphi(x) &=& \frac{W}{%
\varphi^{\gamma+1}} + N\exp(-3x)\exp(-2\varphi),  \notag
\end{eqnarray}
where $x \equiv \log a$, $W \equiv \gamma V_{\ast}/\varsigma^2$ and $N
\equiv 2|\zeta|\rho_{m0}/\varsigma^2$.

\begin{figure}[tbp]
\centering \includegraphics[scale=0.87] {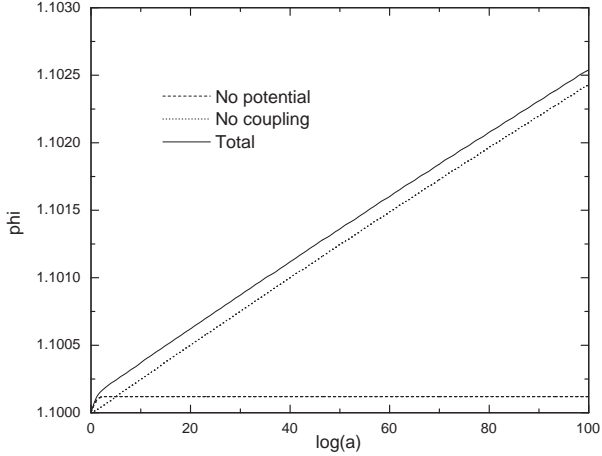}
\caption{The evolution of $\protect\varphi $ as a function of $\log a$ in a $%
\Lambda $-dominated universe. The solid, dashed, and dotted curves represent
the total evolution, the evolution governed solely by the coupling term, and
that governed only by the bare-potential term respectively. The parameters
are $\protect\gamma =50$, $N=W=0.01$, and the initial conditions are $\dot{%
\protect\varphi}_{i}=0,$ and $\protect\varphi _{i}=1.1$.}
\label{fig:Figure6}
\end{figure}

According to the above analysis, when the effective potential $V_{eff}$ is
dominated by the coupling term, $\varphi $ will approach a constant in the $%
\Lambda $-dominated era, while if $V_{eff}$ is dominated by the bare
potential then $\varphi $ evolves as $\varphi \sim t^{\frac{1}{\gamma +2}%
}\propto (\log a)^{\frac{1}{\gamma +2}}$. These qualitative behaviours can
be observed in Fig.~\ref{fig:Figure6}. As $\varphi $ goes large, the bare
potential term in $V_{eff}$ decreases as $\sim \varphi ^{-(\gamma +1)}$
while the coupling term decays as $\sim \exp (-3\varsigma t)\exp (-2\varphi
) $, so eventually the former will always dominate over the latter, driving
the continuous growth of $\varphi $ in contrast to the asymptotically
constant behaviour seen in BSBM where $W=0$.

\begin{table*}[tbp]
\caption{Tracking behaviours of $\protect\alpha $ in different varying-$%
\protect\alpha $ models in different limits. $V$ denotes the bare potential
term and $C$ is the coupling term; $\rightarrow $ means asymptotic approach.}%
\begin{ruledtabular}
\begin{tabular}{cccc}
Model & Radiation Era & Matter Era & Acceleration Era \\ \hline
BSBM & $\alpha\sim t^{\frac{1}{2}}$ & $\alpha \sim \log t$ & $\alpha \rightarrow \mathrm{const.}$ \\
$\mathrm{BSBM} + V_{0}\exp(-\lambda\varphi)$ &
$\alpha\sim\left\{\begin{array}{ll}
t^{\frac{4}{\lambda}}, & \hbox{$V$~dom.} \\
t^{\frac{1}{2}}, & \hbox{$C$~dom.} \\
\end{array}%
\right.$ & $\alpha\sim\left\{
\begin{array}{ll}
t^{\frac{4}{\lambda}}, & \hbox{$V$~dom.} \\
\log t, & \hbox{$C$~dom.} \\
\end{array}%
\right.$ & $\alpha\left\{%
\begin{array}{ll}
\sim t^{\frac{4}{\lambda}}, & \hbox{$a\propto t^{n}$;} \\
\rightarrow t^{\frac{2}{\lambda}}, & \hbox{$H=\mathrm{const.}$.} \\
\end{array}%
\right. $ \\
$\mathrm{BSBM} + V_{\ast}\varphi^{-\gamma}$ & $\alpha\sim\left\{%
\begin{array}{ll}
\exp\left(2t^{\frac{2}{\gamma+2}}\right), & \hbox{$V$~dom.} \\
t^{\frac{1}{2}}, & \hbox{$C$~dom.} \\
\end{array}%
\right.$ & $\alpha\sim\left\{%
\begin{array}{ll}
\exp\left(2t^{\frac{2}{\gamma+2}}\right), & \hbox{$V$~dom.} \\
\log t, & \hbox{$C$~dom.} \\
\end{array}%
\right.$ & $\alpha\left\{%
\begin{array}{ll}
\sim \exp\left(2t^{\frac{2}{\gamma+2}}\right), & \hbox{$a\propto t^{n}$;} \\
\rightarrow \exp\left(2t^{\frac{1}{\gamma+2}}\right), & \hbox{$H=\mathrm{const.}$.} \\
\end{array}%
\right. $ \\
\end{tabular}
\end{ruledtabular}
\label{tab:table1}
\end{table*}

\begin{figure}[tbp]
\centering \includegraphics[scale=1.2] {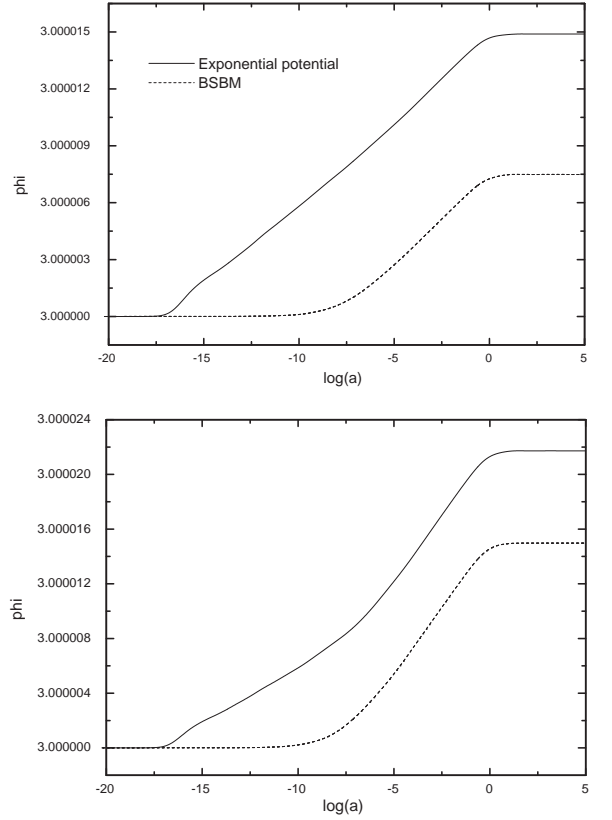}
\caption{Comparison of the entire cosmological evolution of
$\protect\varphi $ as a function of $\log a$ in BSBM model (dashed
curve) and the model with an exponential potential $V=V_{0}\exp
(-\protect\lambda \protect\varphi )$ plus coupling term (solid
curve). The parameters for the upper panel are,
respectively, $\protect\lambda =5\times 10^{6},|\protect\zeta |/\protect%
\omega =10^{-4}$ and $\protect\lambda =5\times 10^{6},|\protect\zeta |/%
\protect\omega =2\times 10^{-4}$. In both cases the initial conditions are $%
\protect\varphi _{i}=3,\dot{\protect\varphi}_{i}=0$.}
\label{fig:Figure7}
\end{figure}

\begin{figure}[tbp]
\centering \includegraphics[scale=1.1] {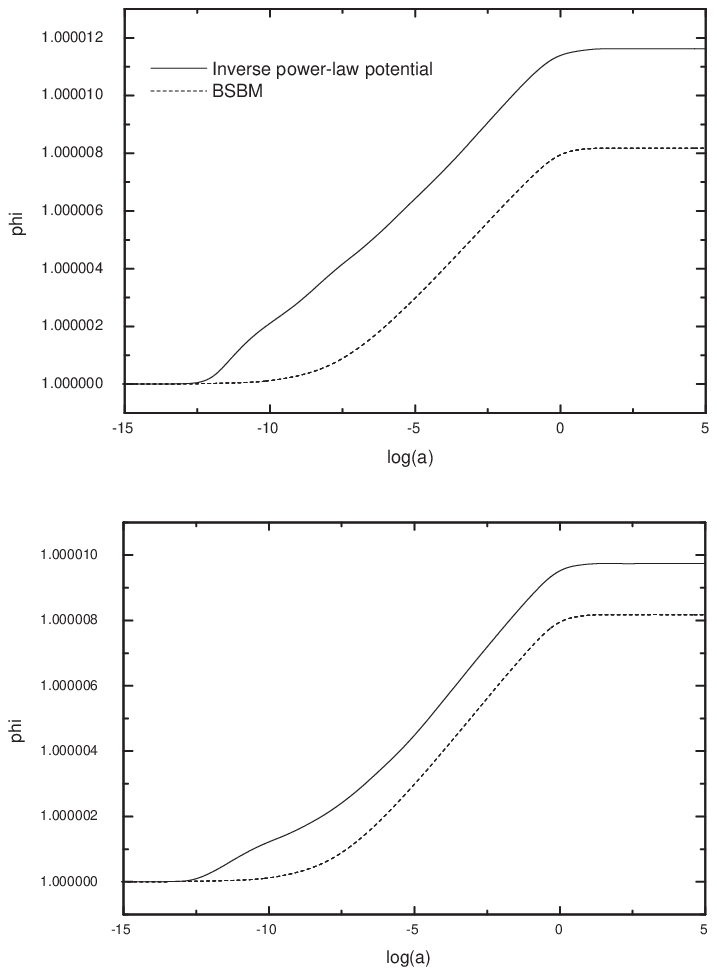}
\caption{Comparison of the entire cosmological evolution of
$\protect\varphi $ as a function of $\log a$ in the BSBM model
(dashed curve) and the model
with an inverse power-law potential $V=V_{\ast }\protect\varphi ^{-\protect%
\gamma \text{ }}$plus coupling term (solid curve). The parameters for the
upper panel are, respectively, $\protect\gamma =5\times 10^{6},|\protect%
\zeta |/\protect\omega =2\times 10^{-6}$ and $\protect\gamma =1\times
10^{7},|\protect\zeta |/\protect\omega =2\times 10^{-6}$. In both cases the
initial conditions are $\protect\varphi _{i}=1,\dot{\protect\varphi}_{i}=0$.}
\label{fig:Figure8}
\end{figure}

\subsection{Summary of the behaviours of $\protect\varphi (t)$ and $\protect%
\alpha (t)$}

We summarise the possible behaviours of $\varphi (t)$ found in the different
varying-$\alpha $ models that we have discussed, in the three different
cosmic eras and different situations (bare potential $V_{eff}$ -dominated or
coupling $\zeta $-dominated) in the Table shown here. The time evolution of $%
\alpha $ is obtained from that of $\varphi $ by using $\alpha \propto \exp
(2\varphi )$. Recall from the figures above that the transitions from
bare-potential domination to coupling domination (or vice versa) are smooth,
so there is a simple pattern for the overall $\varphi $ evolution. However,
note that depending on the initial conditions for $\varphi $, the above
'tracking' solutions may only be reached after a long time (as can be seen
in the figures). In this analysis we have focussed upon extracting the
evolution of $\varphi (t)$, and hence of $\alpha (t)$, but a more detailed
study with a different emphasis could also seek the best-fit observational
parameter set when varying alpha is included, as was done in Ref.~\cite{bass}%
.

\begin{figure}[tbp]
\centering \includegraphics[scale=0.87] {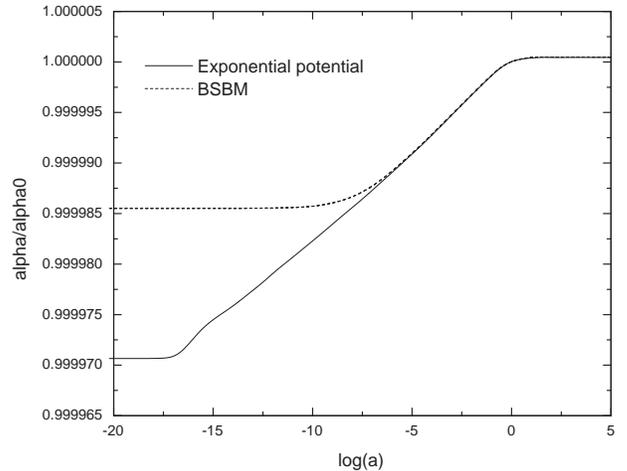}
\caption{The evolution of $\protect\alpha /\protect\alpha _{0}$
versus $\log a$ in the BSBM model (dashed curve) and in the model
with an exponential
potential plus coupling term (solid curve). The parameters chosen are $%
\protect\lambda =5\times 10^{6},|\protect\zeta |/\protect\omega =10^{-4}$.}
\label{fig:Figure9}
\end{figure}

\begin{figure}[tbp]
\centering \includegraphics[scale=0.87] {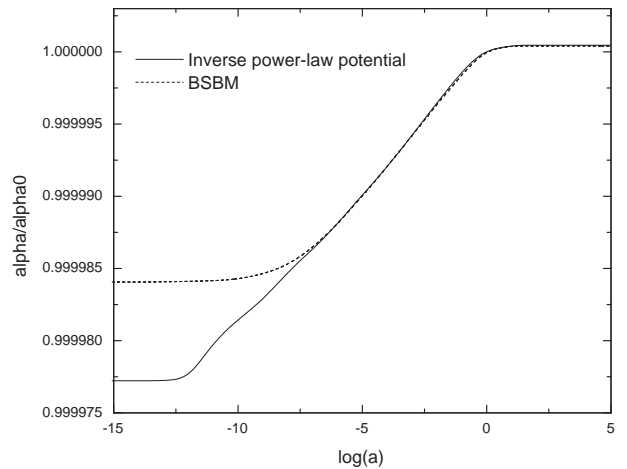}
\caption{The evolution of $\protect\alpha /\protect\alpha _{0}$
versus $\log a$ in the BSBM model (dashed curve) and in the model
with an inverse power-law potential plus coupling term (solid
curve). The parameters chosen are $\protect\gamma =5\times
10^{6},|\protect\zeta |/\protect\omega =2\times 10^{-6}$. }
\label{fig:Figure10}
\end{figure}

\section{Numerical examples of the $\protect\varphi $ and $\protect\alpha $
evolution}

\label{sect:numerics}

We shall now consider the numerical evolution of $\varphi $ and $\alpha $
through the entire cosmological history, and try to connect this evolution
to the observations constraining possible time variation in $\alpha $. As we
have seen above, the evolution in $\varphi $ is controlled by the
competition between the coupling term and bare-potential term in the
effective potential. The parameter $|\zeta |/\omega $ determines the
strength of coupling and so increasing it will increase the rate of
variation of $\varphi $. In addition, in the case of the exponential
potential, $V=V_{0}e\exp (-\lambda \varphi ),$ the parameter $\lambda $
controls the scaling solution of $\varphi $, and the larger it is, the
smaller the fraction of the total energy density $\Omega _{\varphi }$ tends
to be; note also that $\lambda $ governs the slope of the evolution of $%
\varphi $ : if there is no coupling term then the solution of $\alpha
\propto \exp (2\varphi )$ is given by $\alpha \propto a^{-\frac{8}{\lambda }%
} $ in the radiation era and $\alpha \propto a^{-\frac{6}{\lambda }}$ in the
radiation era. This power-law evolution means that in order that the
fractional variation of $\alpha $ between now and $z\sim 6$ should not
exceed the observational bounds, \emph{i.e.}, $\Delta \alpha /\alpha <%
\mathcal{O}(10^{-5}$), $\lambda $ must be very large. Similarly, in the case
of an inverse power-law potential, $\gamma $ controls the scaling solution
of $\varphi ,$ and larger values of $\gamma $ correspond to smaller
variations of $\varphi $ in time, so that an observational constraint that
the allowed variation of $\alpha \propto \exp (2\varphi )$ be small requires
$\gamma $ to be very large.

Finally, the initial value $\varphi _{i}$, is also an important quantity;
although for the exponential potential one can always choose $\varphi _{i}=0$
by adjusting $V_{0}$ correspondingly. For the coupling term we do not have
this freedom -- a larger $\varphi _{i}$ will weaken the coupling through $%
\exp (-2\varphi _{i})$ and thus reduce the change in $\varphi $.

In Figs.~\ref{fig:Figure7} and \ref{fig:Figure8} we have plotted the
evolution of $\varphi (t)$ for some choices of the model parameters for the
two above-mentioned potentials, respectively. For comparison, we also plot
the result for BSBM model (where $V=0$) with the same choice of the
parameter $|\zeta |/\omega $. In both cases $\varphi $ is initially
dominated by the bare-potential term but later becomes dominated by the
coupling term because the largeness of $\lambda $ (or $\gamma $) means that
the slope of the bare-potential-dominated evolution is smaller that the
coupling-dominated evolution at later matter-dominated times. When the
universe becomes dominated by the cosmological constant, the
coupling-dominated solution for $\varphi $ approaches a constant and the
bare-potential-dominated solution grows very slowly, so the total solution
grows very slowly too (almost constant in the figure). Note that the
addition of a bare-potential term makes $\varphi $ begin evolving earlier.
This produces an earlier onset of variation for $\alpha $ than in the pure
BSBM model (the late-time evolution, which is relevant for the QSO
observations, is however almost the same as in BSBM because at this late
stage the total solution is dominated by the coupling term).

Figures~\ref{fig:Figure9} and \ref{fig:Figure10} show the evolution of $%
\alpha /\alpha _{0}$ (where $\alpha _{0}$ is the current value of the fine
structure constant) in the two models when compared with the prediction of
the BSBM model (dashed curves). The qualitative feature in a model with
self-potential for $\varphi $ is the same as in BSBM: at early times $\alpha
$ remains a constant; during the matter-dominated era there is a slow
growth; and then, when the cosmological constant begins to dominate, the
growth stops. The differences are: first, the commencement of growth for $%
\alpha $ starts earlier than the BSBM case because in this model the
bare-potential term could drive the evolution of $\varphi $ even in the
radiation-dominated era; second, the late-time evolution is dominated by the
coupling term and so the late-time evolution of $\varphi $ is similar to
that in BSBM, but the earlier commencement of the growth means that the
total variation of $\alpha $ is greater than that in BSBM. The fractional
change of $\alpha $ between $z\sim 4$ and now is about $0.5\times 10^{-5}$
which is consistent with the QSO observations. The current rate of $\dot{%
\varphi}$ is given by $\dot{\varphi}_{0}=\left( \frac{d\varphi }{dx}\right)
_{0}H_{0}$ where $H_{0}\approx 70~\mathrm{kms}^{-1}\mathrm{Mpc}^{-1}$ is the
present Hubble expansion rate. For the parameters in Figure 12, we have $%
\left( \frac{d\varphi }{dx}\right) _{0}\sim 0.8\times 10^{-6}$ and so we
have $\dot{\varphi}_{0}\sim 0.6\times 10^{-16}/\mathrm{yr}$ which leads to $(%
\dot{\alpha}/\alpha )_{0}\sim 2\dot{\varphi}_{0}\sim 1.2\times 10^{-16}/%
\mathrm{yr}$. This rate is well within all old limits \cite{blatt},\cite%
{fortier},\cite{peiknew},\cite{fischer} but is about an order of magnitude
above the proposed new upper bound \cite{Rosenband} on the current rate of $%
\alpha $ variation from atomic clocks, which is $(\dot{\alpha}/\alpha
)_{0}=(-1.6\pm 2.3)\times 10^{-17}/\mathrm{yr}$, although the uncertainties
may be modulated slightly by accounting for seasonal variations in the local
gravitational potential \cite{bs}.

\section{Summary and conclusions}

\label{sect:conclusion}

In this paper we have considered the dynamics of the varying-$\alpha $
theories which arise when the original BSBM theory is generalised by
introducing an exponential or inverse power-law self-potential for the
scalar field driving the variation of $\alpha $. These two representative
potentials capture the essential ingredients of general potentials without
minima. There are two situations to distinguish and analyse separately:
according as to whether or not the scalar-field potential comes to dominate
the late-time dynamics of the universe. In combination with the coupling
with matter, the additional bare potential forms an effective total
potential $V_{eff}$ for the scalar field $\varphi $ which governs the
allowed time variation of $\alpha $. We have presented the solutions to the
scalar-field equation of motion in cases where $V_{eff}$ is dominated solely
by the coupling term or the bare potential, respectively, in different
cosmic eras. In most cases the bare-potential-dominated solution differs
from the coupling-dominated one; the contributions of these two terms to $%
V_{eff}$ vary with time, and it is possible for there to be a transition
from one solution-type to another. The numerical results show that the
transition between solutions types can be very smooth, and for most of the
time the solution for the scalar field tracks either the
bare-potential-dominated or the coupling-dominated solution. These features
ensure that the evolution of $\varphi $ under $V_{eff}$ has a very simple
pattern. The main results are briefly summarized in Table 1, and these
results are quite general, not depending on whether the parameters defining
the potential ($\lambda $ and $\gamma $) are extremely large or not.

The consequences for the time evolution of the fine structure 'constant' of
adding potentials $V(\varphi )$ \ to the BSBM theory are summarised as
follows. In light of the observational constraints on how much variation in $%
\alpha $ there can be over redshifts $z<6$, we find that the restrictions on
the interaction potential parameters ($\lambda $ in the exponential
potential and $\gamma $ in the inverse power-law potential) must be very
strong, in order to prevent the bare potentials from becoming unacceptably
dominant and driving unacceptable fast time variation of $\alpha $. For
example, in the case of an exponential potential, $V=V_{0}\exp (-\lambda
\varphi ),$ we must have $\lambda \gtrsim 10^{6}\sim 10^{7}$ with $|\zeta
|/\omega \sim \mathcal{O}(10^{-6})$, and for the inverse power-law
potential, $V=V_{\ast }\varphi ^{-\gamma },$ we need $\gamma \gtrsim
10^{6}\sim 10^{7}$ with $|\zeta |/\omega \sim \mathcal{O}(10^{-6})$ -- the
exact constraint, of course, depends also on the initial conditions). This
means that the $\lambda $ and $\gamma $ are constrained to be so large that
if they appear in quintessence models then the scalar field is practically
indistinguishable from a cosmological constant, which has no dynamics at
all. The total variation of $\alpha $ can be enhanced compared to the case
of no bare potential (BSBM), and the variation commences much earlier.
Finally, because with an exponential or inverse power-law potential the
scalar field will not approach a constant even in the dark-energy-dominated
era, the fine structure constant $\alpha $ will continue to increase for all
future time, until eventually there will no stable atoms in the Universe at
all \cite{bsm2}.

%\appendix

\begin{acknowledgments}
BL acknowledges supports from an Overseas Research Studentship, the
Cambridge Overseas Trust, the DAMTP and Queens' College, Cambridge.
\end{acknowledgments}

\end{document}